# Doping effects on charge density instability in non-centrosymmetric $Pb_xTaSe_2$

**A Sharafeev**[1,2], **R Sankar**[3], **A Glamazda**[4], **K-Y Choi**[4], **R Bohle**[1,2], **P Lemmens**[1,2,*] **and F C Chou**[3]

[1]Institute for Condensed Matter Physics, TU Braunschweig, D-38106 Braunschweig, Germany

[2]Laboratory for Emerging Nanometrology, TU Braunschweig, D-38106 Braunschweig, Germany

[3]Center for Condensed Matter Sciences, National Taiwan University, Taipei 10617, Taiwan

[4]Department of Physics, Chung-Ang University, Seoul 156-756, Republic of Korea

*Corresponding Author, E-mail: p.lemmens@tu-bs.de

**Abstract**

We report on the investigation of vibrational and electronic properties of the Pb doped dichalcogenide $Pb_xTaSe_2$ using Raman scattering experiments. We observe a marked variation of the main vibrational modes with Pb concentration $x$. The concentration dependence of the vibrational modes resembles the dependence of the vibrational modes in $TaSe_2$ on the number of crystallographic layers along the c axis direction [1]. The temperature and polarization dependence of Raman spectra of $Pb_xTaSe_2$ revealed additional broad modes in the low frequency regime which are discussed in context of remnant charge density wave, induced disorder, or PbSe phase formed in the interface of Pb and $TaSe_2$ layers.

## 1. Introduction

In searching for novel electronic materials two dimensional (2D), single-layered systems and their electronic states have recently been in the focus of interest. Examples are graphene-like materials based on a single atom species, but also compounds, like chalcogenides. While a structural rigidity and intrinsic large conductivity are necessary preconditions for any application, other properties are also highly relevant. These are the possibilities to introduce charge carriers and a large and non-linear electronic response, which could be due to electronic or electron-lattice coupling. Recently, also the importance of spin-orbit interaction and the control of disorder have been highlighted [2].

Layered chalcogenides show several interesting features which make them promising materials. There exist atomically thin layers with a defined number of lattice constants. Furthermore, doping may be achieved by intercalation. There exist charge density wave instabilities with interesting transport anomalies and comparably high transition temperatures, as well as superconductivity [3].

Upon an intercalation of Pb between the $TaSe_2$ layers, CDW is suppressed and a superconducting transition is increased to $T_c$ = 3.79 K in $PbTaSe_2$ [4]. $PbTaSe_2$ consists of alternating stacking of hexagonal $TaSe_2$ and Pb layers (see figure 1) and constitutes a moderately coupled, type-II BCS superconductor.

Noticeably, the introduction of Pb to the TaSe$_2$ layers induces distinct structural and electronic changes. PbTaSe$_2$ lacks a center of inversion. The broken inversion symmetry, together with strong spin-orbit coupling (SOC) linked to the heavy Pb atom lifts the spin degeneracy of electronic bands and the superconducting states, are given by an admixture of spin-singlet and triplet. The Rashba spin splitting amounts to an order of 0.8 eV, which is comparable to that of the giant Rashba semiconductor BiTeI [5]. Electronic structure calculations reveal that in PbTaSe$_2$ the single-layer Pb sublattice generates a Dirac point at the *K* point of the Brillouin zone. Moreover, it interfaces with the TaSe$_2$ layers to create a superconducting superlattice and 3D massive Dirac fermions [4]. This suggests that PbTaSe$_2$ represents an exceptional system incorporating non-centrosymmetric superconductivity and nontrivial electronic topologies in a single material.

Of great importance is thus to elucidate the role of the Pb layers in determining structural and electronic properties of Pb$_x$TaSe$_2$. Raman spectroscopy is an experimental tool of choice as it is extremely sensitive to local lattice distortions or electronic modulations.

In this paper, we report on detailed Raman scattering measurements of the doped dichalcogenides Pb$_x$TaSe$_2$ with focus on tracing structural and electronic properties as a function of Pb concentration. The key observation is the appearance of new broad peaks in the low frequency regime, which are not part of symmetry-allowed Raman modes. As a possible origin, we discuss the formation of a PbSe phase in the interface between the Pb and TaSe$_2$.

## 2. Experimental details

Single crystals of TaSe$_2$ and Pb$_x$TaSe$_2$ were grown by chemical vapour transport method [6]. Raman scattering measurements were performed on single crystals of different composition from pure TaSe$_2$ to Pb$_x$TaSe$_2$ with $x$ = 0.25, 0.33, 0.5, 0.75, and 1.0 in quasi-backscattering geometry. Light scattering polarizations are given by parallel (*xx*) and cross (*xy*) polarization within the crystallographic *ab* plane. Freshly cleaved sample surfaces were prepared at ambient pressure using scotch tape. Cleaved-off pieces and their opposite faces are quickly cooled down in vacuum to reduce surface deterioration. Temperature was varied between 8 K and 300 K using a closed-cycle cryostat.

As an incident Laser excitation a $\lambda$ = 532 nm solid state laser and an Ar-Kr-ion multiline gas laser ($\lambda$ = 476 nm, 488 nm, 514.5 nm, 568 nm, and 647 nm) were used. The laser power was set to *P*=5 mW with a spot diameter of approximately 100 μm to avoid heating effects and deterioration of the samples which were mounted on a sample holder in vacuum. The six different laser lines allowed to probe effects of resonance Raman scattering.

Experiments at room temperature were performed using a micro-Raman setup (Horiba Labram) with $\lambda$ = 532 nm. This enabled us to assign the symmetry of phonons, $A_{1g}$ (*A'*) and $E_{2g}$ (*E'*) modes, expected for backscattering geometry. The symbols in brackets correspond to Pb$_x$TaSe$_2$. The Raman spectra at low temperatures were collected using a triple Raman spectrometer (Dilor-XY-500) with an attached liquid-nitrogen-cooled CCD (Horiba Jobin-Yvon, Spectrum One CCD-3000V).

## 3. Experimental results

*3.1 Doping dependence of Raman spectra*

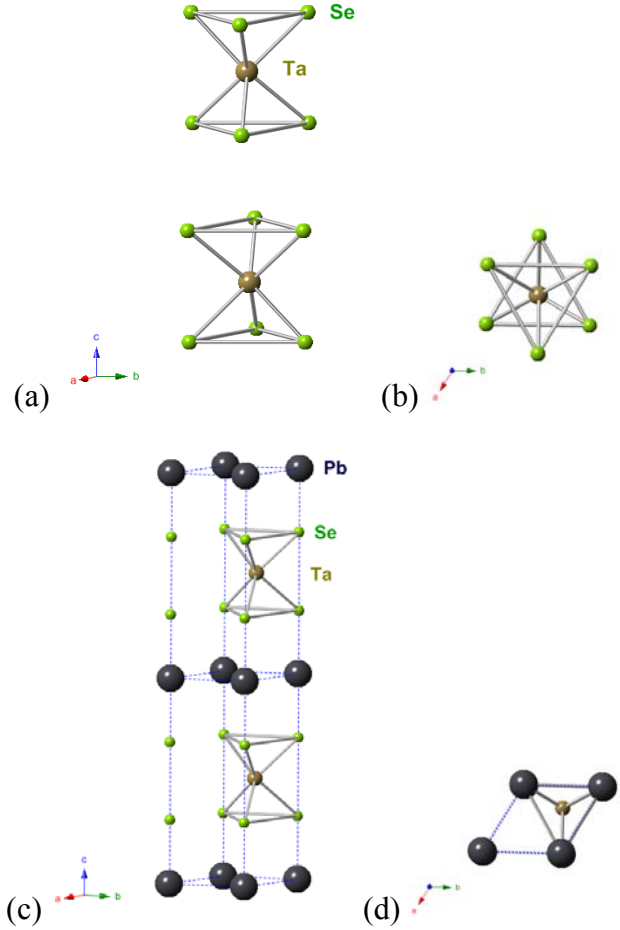

**Figure 1**. (Colour online) Sketches of the crystal structure of a) TaSe$_2$ (hexagonal $D_{6h}^4$ symmetry), b) a view along the c axis and c) PbTaSe$_2$ (trigonal $D_{3h}$ symmetry), and d) a view along the c axis, respectively.

In figure 1, the crystal structures of TaSe$_2$ and PbTaSe$_2$ are depicted. TaSe$_2$ has a layered structure of hexagons that consist of covalently bonded Ta and Se atoms. A plane of Ta atoms is sandwiched between two planes of Se atoms in a trigonal prismatic arrangement. In the high temperature phase, 2H-TaSe$_2$ has a crystal structure of hexagonal $D_{6h}^4$ (space group $P6_3/mmc$) symmetry. For the $P6_3/mmc$ space group the irreducible representation of Raman active modes is given by $\Gamma = A_{1g} + 2E_{2g} + E_{1g}$. We note that $E_{1g}$ modes are not allowed in the measured in-plane polarizations [7].

The crystal structure of Pb$_x$TaSe$_2$ is not yet determined. Rather, for $x = 1$, three possibilities have been proposed as listed in table 1. Trigonal crystal structure $D_{3h}$ (space group $P\overline{6}m2$) is the most probable, which yields $\Gamma = A' + 3E' + E''$ Raman active modes. The $A'$, $E'$ and $E''$ are correlated to $A_{1g}$, $E_{2g}$ and $E_{1g}$ modes of TaSe$_2$, respectively. The Pb doping leads to only one additional mode with $E'$ symmetry.

**Table 1.** Structural parameters and Raman tensors of $TaSe_2$ and $PbTaSe_2$ at room temperature.

| Compound | Space group | c axis (Å) | Ta-Se distance (Å) | Se-Se distance (Å) | Raman tensors, $\Gamma$ | References |
|---|---|---|---|---|---|---|
| $TaSe_2$ | $P6_3/mmc$ ($D_{6h}^4$) | 12.7 | 2.597 | 3.353 | $A_{1g} + 2E_{2g} + E_{1g}$ | [8] |
| $PbTaSe_2$ | $P\bar{6}m2$ ($D_{3h}$) | 9.35 | 2.665 | 3.553 | $A' + 3E' + E''$ | [9] |
| | $P\bar{6}$ ($C_{3h}$) | 9.35 | 2.665 | 3.553 | $A' + 3(^1E') + {}^1E'' + 3(^2E') + {}^2E''$ | [6] |
| | $P6/mmm$ ($D_{6h}$) | 9.35 | 2.665 | 3.553 | $A_{1g} + E_{2g} + E_{1g}$ | [6] |

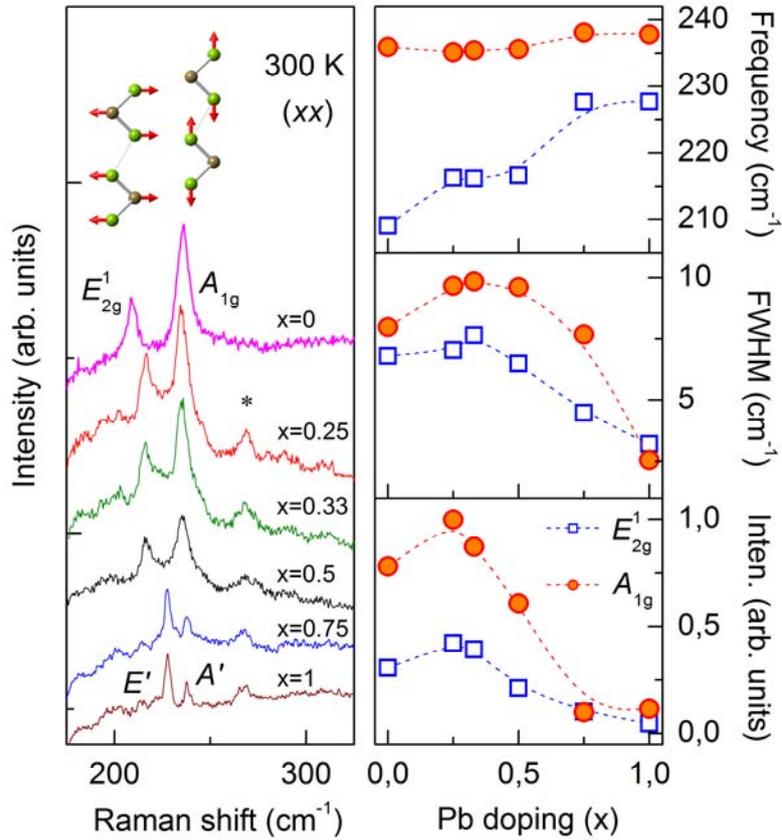

**Figure 2.** (Colour online) (Left panel) Polarized Raman spectra ($xx$) of $Pb_xTaSe_2$ ($x$= 0, 0.25, 0.33, 0.5, 0.75, and 1) measured at room temperature using $\lambda$ = 532 nm. Spectra are shifted vertically for clarity. The insets depict the eigenvectors of the 228 and 238 cm$^{-1}$ modes. The relative amplitude of the vibrations is given by the arrows. The green balls stand for the Se atoms, the olive ones for Ta atoms. The asterisk denotes a new phonon at 267 cm$^{-1}$, which appears upon doping Pb. (Right panel) The frequencies, full widths at half maximum, and integrated intensities of the phonons at 228 and 238 cm$^{-1}$ as a function of Pb doping.

Figure 2 shows Raman data of $Pb_xTaSe_2$ as a function of $x$. These spectra were collected by the micro-Raman setup. Therefore, their frequency range is limited by a filter that cuts off wavenumbers below 130 cm$^{-1}$. The suppression of the low-frequency response is due to an instrument artefact by a notch filter. For $x = 0$, we observe two peaks at 236 and 209 cm$^{-1}$ at $T = 300$ K. Our results are in excellent accordance with data from literature [1]. As depicted in the inset of figure 2, the 209 cm$^{-1}$ mode corresponds to the $E_{2g}^1$ mode involving in-plane stretching motions of the Ta and Se atoms. The 236 cm$^{-1}$ mode is assigned to the out-of-plane vibrations of the Se atoms. We could not detect the anticipated $E_{2g}^2$ mode known as a rigid layer mode due to its very low energy of around 23 cm$^{-1}$ [10].

Upon introducing Pb atoms, a new peak appears at 267 cm$^{-1}$ in addition to the two Raman modes $E_{2g}^1$ (E') and $A_{1g}$ (A') as denoted by the asterisk in the left panel of figure 2. The extra mode is tentatively assigned to 2LO phonon observed in the PbSe nanocrystals [11,12]. The observed phonons are consistent with the factor group prediction for the $P\bar{6}m2$ space group. Increasing the Pb concentration, the phonon modes, pertaining to TaSe$_2$ sublattice, become narrower and are substantially suppressed. The Raman spectra are fitted to a sum of Lorentzian profiles in order to quantify an evolution of the phonon modes as a function of the Pb content. The phonon resulting parameters are plotted in the right panel of figure 2.

With increasing $x$, the in-plane 209 cm$^{-1}$ mode undergoes a large hardening by 19 cm$^{-1}$, suggesting a strong impact of the Pb doping on in-plane electronic properties. A close inspection reveals a step-like variation of the frequency with $x$. This may be related to different Pb ordering phases at low, intermediate and high Pb concentration. Similar effects have been observed in $Na_xCoO_2$ with Na ordering [13,14]. In addition, the drastic suppression of the scattering intensity and a line narrowing occur in the doping range of $x=0.33$. This is ascribed to the screening effects of phonons by electrons in the TaSe$_2$ layer, which become more itinerant through the formation of the Pb layer. This suggests that above $x=0.33$ the interlayer interaction between the PbTaSe$_2$ layers destabilizes CDW, which promotes more metallic behaviour in the TaSe$_2$ plane.

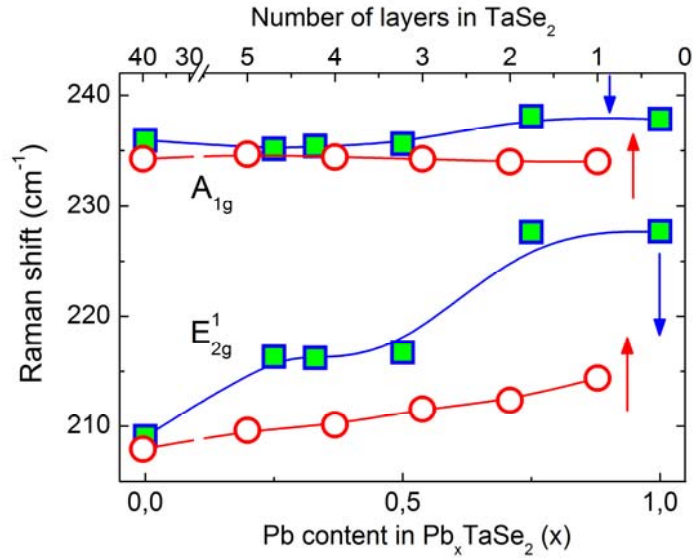

**Figure 3.** (Colour online) Comparison of phonon frequency vs the number of layers (open circles and top axis, respectively) in exfoliated $TaSe_2$ [1] and phonon frequency vs Pb concentration (filled squares and bottom axis, respectively) in bulk single crystals of $Pb_xTaSe_2$.

To analyze further the influence of the Pb layer on the electronic properties, the $x$ dependence of the frequencies of the $E_{2g}^1$ and $A_{1g}$ modes is compared with their frequencies as a function of the number of layers corresponding to the thickness of exfoliated crystals in figure 3 [1]. Overall, the $A_{1g}$ mode varies little with $x$ and thickness. The observed phonon shift is an order of ~1 cm$^{-1}$. This is contrasted by the $E_{2g}^1$ mode, which hardens by ~19 cm$^{-1}$ with increasing $x$ for $Pb_xTaSe_2$ and by 5 cm$^{-1}$ with increasing number of layers in the exfoliated crystals [1]. Here we stress that the Pb doping exerts a much stronger impact on the electric properties than the thickness variation. Obviously, this is linked to the interfacial effect between the graphene-like Pb and the superconducting $TaSe_2$ layers, which is absent for the exfoliated sample $TaSe_2$.

## 3.2 Temperature dependence of Raman spectra

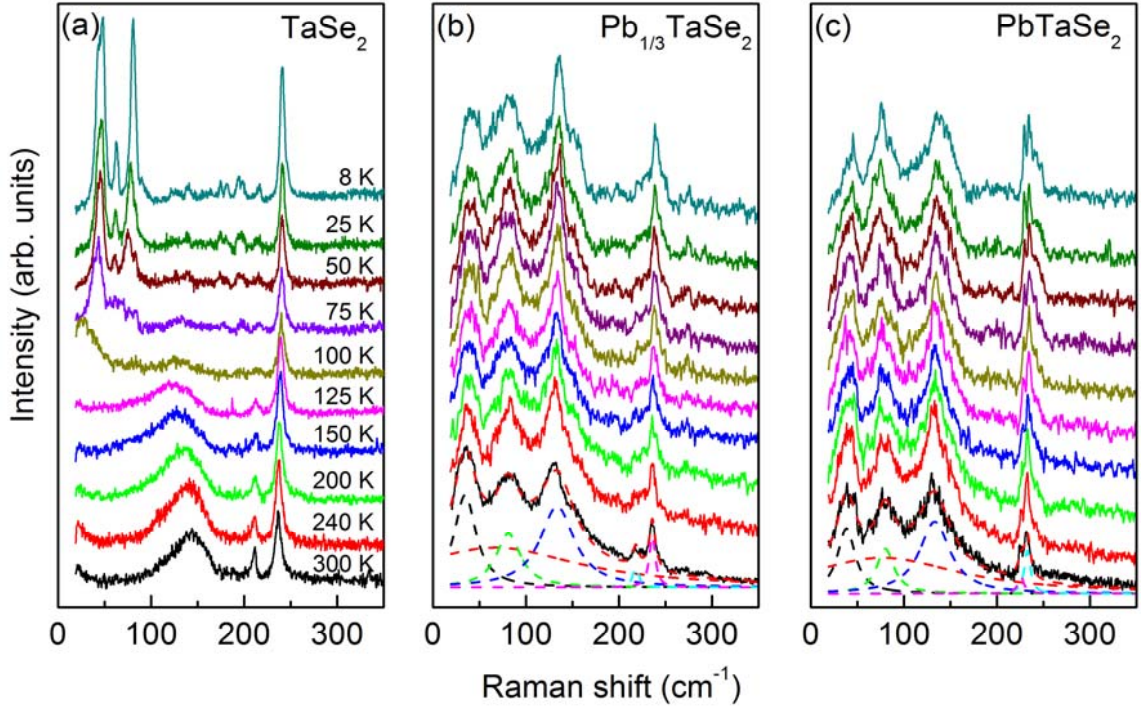

**Figure 4.** (Colour online) Polarized Raman spectra of TaSe$_2$ (a), Pb$_{1/3}$TaSe$_2$ (b) and PbTaSe$_2$ (c) in (*xx*) polarization with different temperatures from 300 K to 8 K. The Raman spectra are offset for clarity. The dashed lines represent a fit of phonon modes together with a background.

We now turn to the temperature dependence of Raman spectra of Pb$_x$TaSe$_2$ (*x*=0, 0.33, and 1), shown in figure 4. For temperatures T> T$_{ICDW}$ = 123 K, TaSe$_2$ exhibits a broad maximum at around ~ 140 cm$^{-1}$ in addition to the two phonon modes discussed above. The former has been attributed to two-phonon Raman scattering amplified by anharmonicity in systems with a Kohn anomaly [15,16]. In the temperature range between T$_{ICDW}$ and T$_{CCDW}$ the two-phonon scattering evolves to a quasielastic response. Below T$_{CCDW}$ = 90 K a number of sharp, well-defined Raman modes at 46, 63, and 81 cm$^{-1}$ appear. They are linked to the CDW 3*a* × 3*a* × *c* superlattice [15,16]. As to Pb$_x$TaSe$_2$ (*x*=0.33 and 1), the broad maxima at 40, 80, and 135 cm$^{-1}$ are present in the whole measured temperature range while the background is being suppressed with decreasing temperature (see figures 4(b) and 4(c)). Their assignment will be made below.

In figure 5, we summarize the temperature dependence of the frequency, the linewidth and intensity for TaSe$_2$. Both $E_{2g}^1$ and A$_{1g}$ modes show a distinct change through T$_{CCDW}$ and T$_{ICDW}$. In particular, the linewidth appreciably decreases and the intensity increases steeply upon cooling through T$_{CCDW}$. This is due to the partial depletion of electronic states at a Fermi surface, which leads to a weakening of charge screening effects.

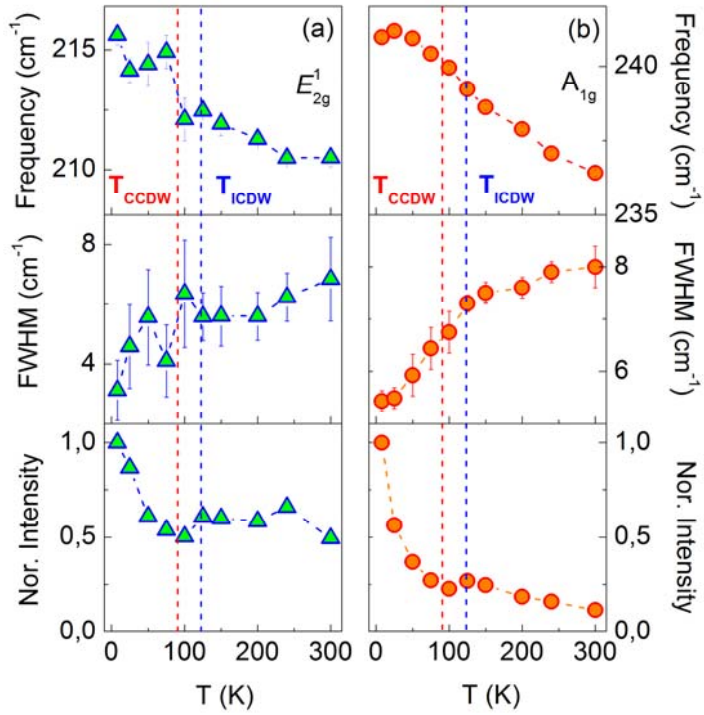

**Figure 5.** (Colour online) Temperature dependence of the frequency, the linewidth and intensity of TaSe$_2$ for (a) $E_{2g}^1$, and (b) for $A_{1g}$ intrinsic modes, respectively. The dashed lines denote the incommensurate CDW and commensurate phase transition temperatures $T_{ICDW}$ = 123 K and $T_{CCDW}$ = 90 K, respectively.

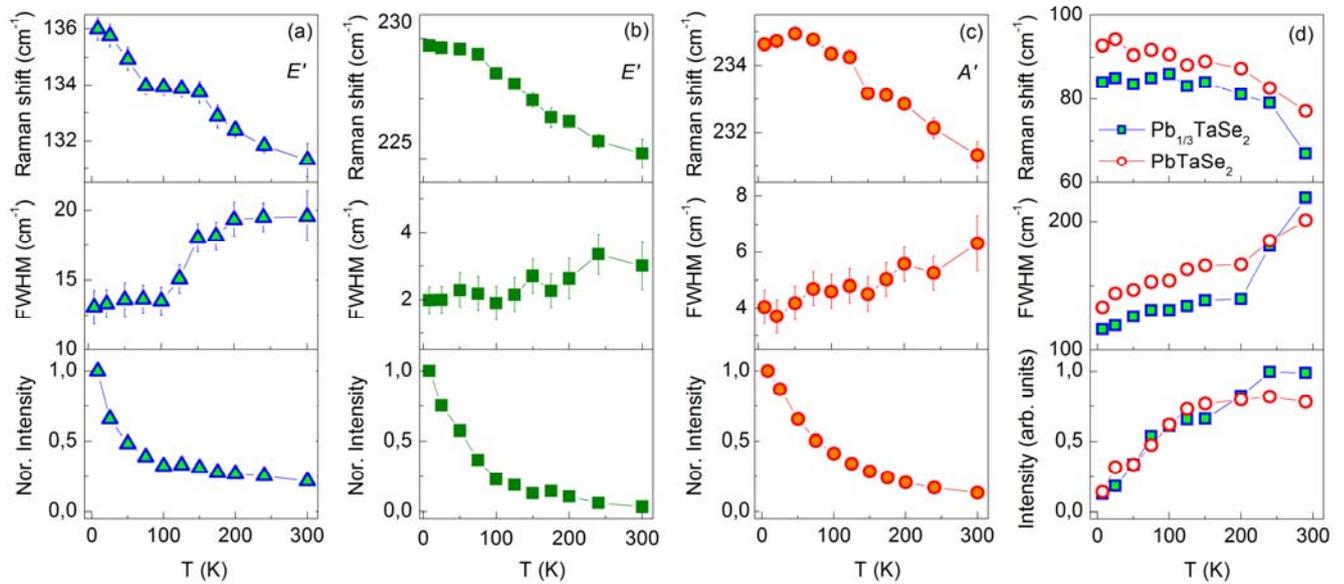

**Figure 6.** (Colour online) Temperature dependence of the frequency, the linewidth and intensity of PbTaSe$_2$ for (a) $E'$ (Pb-Pb) mode, (b) $E'$ mode (c) $A'$ mode and (d) the background signal. The phonon intensities were corrected by the Bose factor but the background intensity consists of raw data.

Next, we will discuss the temperature dependence of the phonon modes and the background scattering for PbTaSe$_2$, which is summarized in figure 6. The phonon frequency hardens by 3-5 cm$^{-1}$ with lowering temperature. This is explained by anharmonicities in the lattice potential energy as commonly observed, e.g. in MoS$_2$ and graphene [17]. This is comparable to the change in frequency, $\Delta\omega = 4.5$ cm$^{-1}$, of TaSe$_2$. The FWHM of the in-plane $E'$ and $A'$ mode of PbTaSe$_2$ is smaller than the respective mode of TaSe$_2$ and the change of the FWHM over a temperature range of T=8-300 K amounts to $\Delta\Gamma \approx 1$-2 cm$^{-1}$ for PbTaSe$_2$ and $\Delta\Gamma \approx 5.7$ cm$^{-1}$ for TaSe$_2$. No discernible anomaly can be found in the temperature dependence of the FWHM. In contrast, the $E'$ (Pb-Pb) mode exhibits a broad signal and a step-like decrease of the FWHM between 100 and 150 K (see figure 6(a)). At the respective temperature, the integrated intensity increases steeply. The same trend is visible for the $E'$ and $A'$ mode as well. Furthermore, we would like to draw attention to the temperature dependence of the background scattering. As evident from figure 6(d), the background signal decreases substantially for temperatures below 150 K. Most probably, the background response originates from light scatterings of electrons by Pb defects.

Combining the above finding together, it seems that the charge dynamics of the hybrid Pb and TaSe$_2$ layers is different from that of the pure TaSe$_2$ layers. Topologically protected materials show coherent surface states at low temperatures whereas incoherent dynamics becomes dominant due to thermally excited phonons at high temperatures [18]. In our case, phonons mediate the Pb and TaSe$_2$ layers, thereby changing the charge dynamics upon heating. This may explain the appearance of the background scattering by Pb defects and phonons at high temperatures and the phonon anomalies of the $E'$ (Pb-Pb) mode.

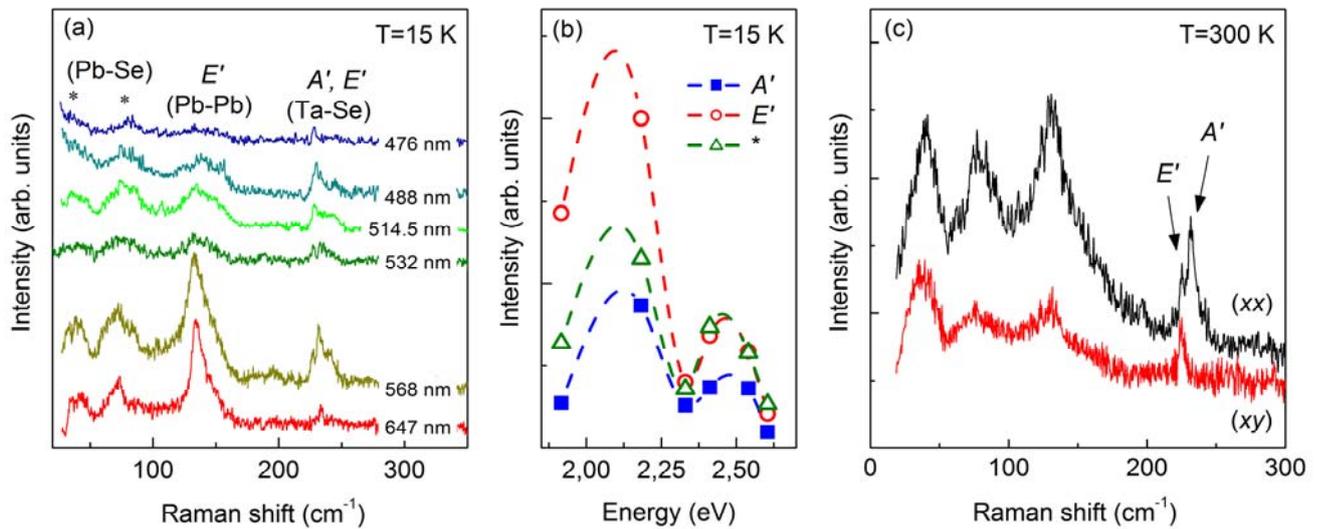

**Figure 7.** (Colour online) (a) Raman spectra of PbTaSe$_2$ with different incident laser excitations measured in (xx) polarization and at T = 15 K. (b) Laser wavelength dependence of the intensity of the E', A' and * modes. The asterisks denote phonon modes related to PbSe nanocrystals. The symbols denote experimental points and the broken curves represent B-spline fit drawn to guide the eye. (c) Polarization dependence of Raman spectra of PbTaSe$_2$ at T = 300 K in (xx) and (xy) polarizations.

We performed resonance Raman scattering measurements on PbTaSe$_2$ with six different laser wavelengths. Respective data and analysis of the intensity are shown in figures 7(a) and 7(b). A resonance effect of the Raman intensity is expected if the incident photon energy matches the energy of respective electronic transitions. In PbTaSe$_2$ the phonon modes indeed show a strong resonance effect in intensity with a maximum at 568 nm (2.18 eV). A sharp minimum is observed at 647 nm (1.92 eV). The largest magnitude of the resonance effect has the $E'$ (Pb-Pb mode) line, for instance $I_{568}/I_{532} \approx 5$, whereas for the $A'$ mode and for the * modes this intensity ratio is $I_{568}/I_{532} \approx 3.3$. Lone-pair $p$ electron states of Pb with their high polarizability are a possible reason for this enhancement.

In figure 7(c) polarization dependent Raman spectra of PbTaSe$_2$ are shown. The $A'$ out-of-plane vibrational mode is suppressed in crossed ($xy$) polarization, while the $E'$ in-plane vibrational mode remains unaltered as expected from the selection rules. The broad mode at ~40 cm$^{-1}$ exhibits little polarization dependence, whereas the other two broad modes around 80 and 130 cm$^{-1}$ show sizable changes. However, we cannot identify apparent selection rules of the low-energy broad peaks. This indicates that they are related to disorders or defects induced by Pb intercalation. This effect will be described in details further below.

## 4. Discussion

In the first part of the discussion the relation of properties such as thickness of TaSe$_2$ layered compound, Pb concentration in Pb$_x$TaSe$_2$ and the covalency of the Ta-Se bond will be considered and related to interlayer and intralayer interactions and their effect on the phonon frequencies. In the second part we consider different possible origins for the low-frequency broad modes, which appear upon Pb intercalation into TaSe$_2$. We assume that such phenomena could be related to the remnants of CDW phase in the Pb doped compounds or induced disorder and defects or the formation of PbSe nanocrystals.

For atomically thin, exfoliated layers of TaSe$_2$ and MoS$_2$ a 3% hardening of the $E_{2g}$ modes and a very weak softening of the $A_{1g}$ have been observed with decreasing thickness. The latter corresponds to a decreasing number of unit cells [1,19]. In a classical model of coupled harmonic oscillators, however, a hardening of phonon modes is expected only with increasing coupling. The opposite effect in the experiments points to an additional interaction which is supposed to be a long range Coulomb interaction from a charge transfer between the layers.

This apparent contradiction also exists for Pb$_x$TaSe$_2$ with an even more pronounced 10% hardening of the $E_{2g}$ phonon with increasing Pb concentration. The geometric changes of Pb$_x$TaSe$_2$ with increasing x can be summarized as an increase of the layer distance, its thickness, the distance between Ta and Se atoms and the Se-Ta-Se angle. We assume that they reflect a change in the electronic configuration and bonding induced by Pb intercalation and, in their turn, induced charge in the layer. The formal oxidation states of Ta and Se in TaSe$_2$ is +4 and -2. With full Pb intercalation the Ta valence changes to +2. This reduction reduces ionic interactions and increases covalent bonding contributions. The Ta-Se interaction shows bigger effect, while the Se-Se mode is largely unchanged. The insensitivity of the $A_{1g}$ mode to the Pb intercalation may be due to the fact that van der Waals (VdW) and long range Coulomb interactions are compensated by charge dynamics. It is interesting to note that such interplay of long range Coulomb interactions between chalcogen atoms and metals together with VdW forces are also discussed for GaS [20] and GaSe [21].

In the following we will focus on doped $Pb_xTaSe_2$ samples and broad modes at low frequencies. Firstly, we propose that after intercalation of Pb layer between Ta-Se layers, the CDW modes could remain afterwards. In figure 4 one can see that the frequencies of the two highly intense CDW modes from $TaSe_2$ are similar to the ones in $Pb_xTaSe_2$ (at about 40 and 80 $cm^{-1}$). Nevertheless, resistivity measurements [4] unveil the absence of CDW in the compound. Similar investigations were done also for copper intercalated superconductor $Cu_xTiSe_2$. Titanium diselenide is a CDW material with $T_{cdw}$ = 200K and upon Cu doping this compound becomes superconducting with $T_c$ = 4.15 K [22]. A Raman scattering study shows that the $E_g$ and $A_{1g}$ CDW amplitude modes are heavily suppressed with concentration of Cu of 5% and the CDW phase is not present anymore.

Albeit, there are cases where the phase transition temperature $T_c$ of a CDW state increases with certain doping. This is the case for $IrTe_{2-x}Se_x$ [23]. Also, it is worth to mention that the change of the Ta oxidation state to +2 introducing two electrons to $Pb_xTaSe_2$ may also lead to some sort of phase separation or charge modulation. Such a short range order assumes a partial localization of these states that are not available for transport. Further studies are needed to clarify this situation and to evaluate whether these states play a role in enhancing the superconducting transition temperature.

In this paragraph we discuss arguments against a residual CDW in the doped samples. We focus on the broad feature around ~130 $cm^{-1}$ consisting of a broad mode and superimposed sharp mode (Pb-Pb interaction [24]). Very likely, the broad mode could be assigned to a second-order Raman process since a similar peak is observed in $TaSe_2$. In accordance to Raman measurements of transition metal dichalogenides, the 2-phonon Raman mode has different temperature dependence in comparison with $PbTaSe_2$ [15]. Normally, 2-phonon Raman scattering is strongly suppressed with decreasing temperatures and the peak shifts towards lower energies. This process is distinctive for CDW-type materials, such as $TaSe_2$ and $NbSe_2$ [25]. Nevertheless, in $PbTaSe_2$ the behaviour of the broad mode is different compared to the usual 2-phonon mode in transition metal dichalcogenides. In our case, the broad mode at ~130 $cm^{-1}$ in $PbTaSe_2$ is not suppressed at low temperatures. Such behaviour is incompatible with a CDW scenario.

A second possible explanation for the broad modes at 40, 80 and 135 $cm^{-1}$ is related to a structural transition due to intercalation. Upon Pb doping, the crystal structure changes drastically. For instance, the layers of Ta and Se atoms in $TaSe_2$ form a "star" along the $c$ axis and have an inversion centre (see figure 1). However, after intercalation of Pb atoms into the structure, such the formation is triangular, leading to a non-centrosymmetric system with no more inversion centre. Due to this process, besides Ta-Se Raman modes, one could observe additional modes possibly from Pb-Pb or/and Pb-Se interaction.

There is another perspective related to a possible formation of Pb clusters. Spiro *et al.* [24] performed Raman measurements on $Pb_6O(OH)_6^{4+}$ containing Pb clusters. They observed several modes related to Pb-Pb interactions. The Raman frequency of such interactions depends on the distance between Pb atoms. A distance between Pb atoms of 3.44 Å has been attributed with an intense and sharp peak at 150 $cm^{-1}$ in $Pb_6O(OH)_6^{4+}$. In this respect, the 130 $cm^{-1}$ mode observed in $PbTaSe_2$ may be due to Pb-Pb clustering. As Pb atoms can have different positions in the clusters with a variation of distances, other broad modes could have the same origin. In $Pb_6O(OH)_6^{4+}$ there also exist broader modes at low frequencies that superimpose sharper excitations [24].

The third reason for the appearance of these broad modes is provided by the formation of PbSe nanocrystals or Pb-Se bondings. The polarized Raman scattering experiments showed that the A' and E' vibrational modes follow the selection rules (figure 7). However, the additional broad modes have little difference between the (*xx*) and (*xy*) polarization. When Pb atoms are inserted between the $TaSe_2$ layers, predominately the Pb atoms form a Pb layer by Pb-Pb bonding. Nonetheless, it is possible that they form PbSe nanocrystals through Pb-Se bonding. Indeed, lead chalcogenides Pb*X* (*X* = S, Se, or Te) are stable semiconductors and their Raman spectra are scattered between 30 and 130 $cm^{-1}$ with second-order scattering observed at about 260 $cm^{-1}$ [11,12]. This accounts for the first-order peaks at 40, 80 and 135 $cm^{-1}$ as well as the second-order peak 267 $cm^{-1}$, which are denoted by the asterisks in figures 2 and 7. Such a scenario provides an explanation for their obscure polarization and temperature dependence distinctly different from the E' (Pb-Pb) mode, E' and A' modes. Therefore, we conclude that doping $TaSe_2$ with Pb atoms forms a heterogeneous Pb-Se bonding in addition to the Pb layer.

## 5. Conclusions

To conclude, we have presented extensive Raman scattering experiments of $Pb_xTaSe_2$. The concentration dependence of Raman spectra showed a pronounced hardening of the in-plane $E_{2g}$ mode with increasing Pb content, which is much larger than an effect observed in exfoliated crystals with different layer thickness. This can be taken as a signature of nontrivial coupling between Pb and $TaSe_2$ layers, which is responsible for structural and electronic changes. In addition, for the Pb doped compounds the temperature and polarization dependence of Raman spectra reveal three additional broad modes in the low frequency range, which behave differently from the symmetry-allowed Raman modes. Their origin is discussed in terms of PbSe nanocrystals or clusters formed at the interface between the Pb and $TaSe_2$ layers.

**Acknowledgement**

For important discussion we thank V. Gnezdilov, Yu. G. Pashkevich, and A. Möller. We also appreciate the support of the *International Graduate School of Metrology* (B-IGSM) and the Graduate School *Contacs in Nanosystems*. Furthermore, financial support by GIF and DFG-RTG 1952 *Nanomet* is acknowledged. KYC is thankful for the support from the National Research Foundation of Korea (NRF) (Grants No. 2009-0093817 and No. 2012-046138).